\documentclass[epj]{svjour}
\usepackage{graphics}
\usepackage{notoccite}
\usepackage[table,xcdraw]{xcolor}
\usepackage{amsmath}
\begin{document}
\title{Production cross sections of tetraquark states in elementary hadronic collisions}
\author{Gábor Balassa\inst{1} \and György Wolf \inst{1} 
}                     
%
%
\institute{\inst{1}Institute for Particle and Nuclear Physics, Wigner Research Centre for Physics, H-1525 Budapest, Hungary}
\date{Received: date / Revised version: date}
%
\abstract{
Inclusive production cross sections of the possible exotic state $X(3872)$ in proton-proton, pion-proton and proton-antiproton collisions are calculated using a statistical based model, which is previously used to describe inclusive charmed and bottomed hadron production cross sections in the low energy region. With the extensions made here the model is capable to include tetraquarks as well, using the diquark picture of tetraquarks. The evaluated cross section ratio of $\Psi(2S)$ and $X(3872)$ at $\sqrt{s}=7$ TeV agrees well with the measured value.
\PACS{
      {PACS-key}{discribing text of that key}   \and
      {PACS-key}{discribing text of that key}
     } 
} 
\maketitle
\section{Introduction}
\label{intro}
The possibility of exotic states are an interesting and still on-going research topic, as nowdays more and more experiments aim to produce and measure these particles, with great accuracy. These exotic states like tetra\-quarks, pentaquarks, glueballs etc.. are long time predicted by quantum chromodynamics, however their status are still not fully established yet \cite{Tetraquark1,Tetraquark3,Tetraquark5,Tetraquark6,Tetraquark7,Tetraquark8}. To make conclusions about the nature of a specific exotic state, one way is to use an appropriate model to calculate its properties e.g. masses, decay widths. and compare it to experiments. Such a model is for example the bag model \cite{BAG1,BAG2}, which is able to give good estimates to the hadron masses, and can be used to calculate e.g. the masses and the excitation spectrum of e.g. the glueball states as well \cite{BAG_GLUEBALL}. It is also capable to describe the decay widths with a string breaking mechanism described in \cite{String_model,StringBreaking}, where that specific model is also used to explain the possible tetraquark excitation spectrum \cite{String_model_TETRA}. Other models like non-relativistic quantum chromodynamics (NRQCD) can be also used to describe the properties of states containing heavy quarks e.g. hidden charmed tetraquarks \cite{Tetraquark_NRQCD}. Solving the many-body non-relativistic Schrödinger equation is also a possibility to estimate the mass spectrum of tetraquarks and pentaquarks containing heavy quarks \cite{TetraquarkSCH}. Effective field theoretical methods \cite{Tetra_EFT}, lattice quantum chromodynamics \cite{Tetra_LQCD}, and QCD sum rules \cite{TetraquarkSUM} are also tools to understand these states better. All of these methods are relying on some free parameters e.g. coupling constants, quark masses, which have to be tuned to existing data to be able to give reliable estimates, so it is necessary to make more and more measurements and compare the different approaches to each other to fully understand these states.
In this paper, we examine the possible hidden charmed tetraquark state $X(3872)$, which is measured at the Belle collaboration in 2003 from $B \rightarrow X K^+$ and $X \rightarrow J/\Psi \pi^+ \pi^-$ decays \cite{BELLE}. Its quantum numbers are expected to be $J^{PC}=1^{++}$ and due to its significant $\approx 100$ MeV mass difference from the $c\overline{c}$ model it is theorized to be an exotic state with a quark content of $(c\overline{c}u\overline{u})$ . Its actual form is however still unknown, as it could be a loosely bound  $D^0 \overline{D^{*0}}$ meson-meson molecule, a diquark-antidiquark bound state, a compact 4 quark state, a charmonium hybrid $c\overline{c}g$ or a glueball mixing with charmonium states as well \cite{TetraPentaquarks}. The average size of each state and therefore their production cross sections and reaction rates are highly dependent on the model used, which could be a good tool to make decisions about their actual structure e.g. in heavy ion collisions \cite{X_in_Medium,X_TRANSPORT1,X_TRANSPORT2}. However to make estimations with the help of heavy ion tranport models the necessary ingredients are the elementary cross sections to create tetraquarks, which are not known, as only a few measurement points are available so far. There are possibilities to estimate them however. In \cite{X_TRANSPORT3} effective model calculations are carried out to estimate the $X(3872)$ production rate, while in \cite{TetraquarkDPS1} the double partonic scattering model is used to describe the high energy cross sections at the TeV energies. Here, we propose a different, statistical based approach to calculate inclusive and exclusive hardonic cross sections with tetraquark states in the final states, using the diquark-antidiquark picture.

In the diquark picture two quarks are bound together by the strong force to create a diquark. Because the quarks are in the color SU(3) fundamental representation the diquark could be in a antitriplet or a sextet color state according to the tensor product $3 \otimes 3 = \overline{3} \oplus 6$. For antiquarks the process is the same with the two possible antidiquark color state $3$ and $\overline{6}$. The diquarks and antidiquarks then could create a color singlet state according to $\overline{3} \otimes 3 = 1 \oplus 8$ and $6 \otimes \overline{6} = 27 \oplus 8 \oplus 1$. It is not straightforward however to estimate the mixing between the triplet and sextet states and therefore it is mostly assumed that the diquark is in the triplet state \cite{Tetraquark4}.

In Sec.\ref{sec:2} we briefly summarize the statistical model, which is used to estimate tetraquark production rates, highlighting the extra contributions due to the diquark picture, we apply here. In Sec.\ref{sec:3} we introduce the necessary estimations to a necessary ingredient - the quark creational probabilities - at very high energies in the TeV range, where due to the limited data a simple fit was not possible, as it was before at lower energies. This estimation is needed to be able to validate the model using the measured value at 7 TeV CM energy in proton-proton collisions. In Sec.\ref{sec:4} the results for the inclusive direct production cross section calculations of the $X(3872)$ particle in proton-proton, pion-proton, and antiproton-proton collisions are shown, while Sec.\ref{sec:5} concludes the paper.

\section{Statistical model approach}
\label{sec:2}
In this section, we introduce and briefly summarise a statistical model developed by us to estimate inclusive tetra\-quark cross sections in the diquark picture. The model is introduced in \cite{Main1} where it was used to describe low energy exclusive cross sections and proton-antiproton annihilation at rest. In \cite{Main2} the model is extended to describe inclusive cross sections at higher energies as well, while in \cite{Main3} it is used to describe inclusive charmonium, bottomonium and open charm cross sections as well. In this work it is extended further to describe tetraquarks in the diquark picture which is the most straightforward way to include these states into the model.
The model starts from the assumption that during a hadronic collision a so called fireball is formed, which will hadronize into some specific hadronic final state with a probability described by the model. In its most general form the transitional probability is factorized into a dynamical part, which describes the initial dynamical part of the collision, and a statistical part describing the hadronization to a final state as it is shown in Eq.\ref{eq:25}

\begin{eqnarray}
\label{eq:25}
\nonumber
&{\displaystyle \sigma^{n \rightarrow k}(E)=\Bigg( \int \prod_{i=1}^{n} d^3p_i R(E,p_1,...,p_n) \Bigg) } \\
&{\displaystyle \times \Bigg( \int \prod_{i=1}^{k} d^3q_i w(E,q_1,...,q_k) \Bigg)}
\end{eqnarray}
where $ \sigma^{n \rightarrow k}$ is a generalized transitional probability for an $n \rightarrow k$ process. The function $R(E,p_1,...p_n)$ describes the initial stage of the collision, which depends on the momenta of the colliding particles $p_i$ and the center of mass energy $E$, while $w(E,q_1,...,q_k)$ describes the hadronization of the fireball, which depends on the momenta of the outgoing particles $q_j$ and the center of mass energy. Hereafter, we only consider two-body collisions and make the assumption that the dynamical part is described by the inclusive cross section of the two-body collision:
\begin{equation}
\sigma_{I}^{1+2 \rightarrow X} \equiv \int  d^3p_1 d^3p_2 R(E,p_1,p_2)
\label{eq:26}
\end{equation}
where $\sigma_{I}^{1+2 \rightarrow X}$ is the inclusive cross section of the two-body process. The statistical part which gives the probabilities of a specific many body final state e.g. $p p \pi^+$ consists 4 main ingredients: (1) the probability of a specific fireball configuration, (2) density of states from the statistical bootstrap approach,  (3) phase space integrals including Breit-Wigner factors for resonances, (4) quark creation probabilities and quark combinatorial factors. Here a short description is given for each factor, while a detailed description is available in \cite{Main1,Main2,Main3}.

After the initial stage the created fireball is modeled as an object, which could decay into more, smaller fireballs as follows. At the first step a fireball with invariant mass $M$ is formed which can decay into two smaller fireballs, which probability is determined by using the assumption that the decay follows an uniform distribution in energy, and constrained from below by the masses of the neutral pions. The sum of the masses of the daughter fireballs has to be equal to the parent fireball, which in this case $m_{11}+m_{12}=M$. After the two fireball is formed each of them could decay into more fireballs or hadronize e.g. if $m_{11}$ decides to decay to another two fireballs but $m_{12}$ wants to hadronize then we will have $m_{11} \rightarrow m_{21},m_{22}$ and $m_{12}$, with the constraint $M=m_{21}+m_{22}+m_{12}$. The decay chain then continues until all the fireballs want to hadronize. At the end there will be an $n$ fireball constuction with the probability $P_n^{fb}(E)$, which is explained in all its detail in \cite{Main1}. After a specific $n$-fireball structure is formed it will start to decay into hadrons. At the first step, we assume that the number of hadrons coming from one fireball is two or three with probabilities $P_2^d=0.69$ and $P_3^d=0.24$, which is a result coming from the statistical bootstrap approach \cite{Bootstrap1,Bootstrap2,Bootstrap3,Bootstrap4,Bootstrap5}. There is a small probability for more than three hadrons, however it is negligible. In this way the minimum number of final state hadrons from an $n$-fireball construction are $2n$, while the maximum number of hadrons are $3n$. The probability of a specific final state is then given by phase space factors, quark combinatorial factors etc.. which will be described below.

In the most general form the hadronization probability of $k$-fireballs is given by Eq.\ref{eq:27}.
\begin{eqnarray}
\label{eq:27}
\nonumber
&{\displaystyle W_{k,i_1..i_k}(E) =P_k^{fb}(E)  \frac{1}{Z_k}\frac{1}{N_{i_1,..i_k}!}\int_{x_{min}}^{x_{max}} \prod_{a=1}^{k} } \Bigg[ dx_a  \times \\ 
& {\displaystyle\times \frac{T_{i_a}(x_a)}{\sum_j T_j(x_a)}  \delta \Big( \sum_{a=1}^k x_a-E \Big)  \Bigg]},
\end{eqnarray}
where $P_k^{fb}(E)$ is the probability of the $k$-fireball scheme, $N_{i_1,..i_k}$ is the number of fireballs with the same hadrons in their final states, $T_{i_a}(x_a)$ is a factor consisting the remaining phase space integrals, quark combinatorial factors and density of states, while $Z_k$ is an overall normalization function given by:
\begin{eqnarray}
\label{eq:28}
&{\displaystyle Z_k(E)=  \sum_{<l_1..l_k> \in S} \frac{1}{N_{l_1,..l_k}!} \int_{x_{min}}^{x_{max}} \prod_{a=1}^{k} } \Bigg[ dx_a  \times \\ \nonumber
& {\displaystyle\times \frac{T_{l_a}(x_a)}{\sum_j T_j(x_a)}  \delta \Big( \sum_{a=1}^k x_a-E \Big)  \Bigg]},
\end{eqnarray}
where the summation goes for all of the possible processes, which have the correct quantum numbers $S$=(baryon number, electric charge, strangeness, charmeness, and bottomness).

For one fireball the factor $T(x)=C_{Q}(x)P_{n}^{H}(x)$ \footnote{We have dropped the fireball index $i_a$ for simplicity and use the letter $x$ for the energy of the fireball. } contains the quark combinatorial factors, and phase space integrals, where $C_{Q}(x)$ is the quark combinatorial factor, and $P_{n}^{H}(x)$ is given by:
\begin{equation}
\label{eq:29}
P_{n}^{H}(x) = P_n^d \frac{\Phi_n(x,m_1,..,m_n)}{\rho(x)(2\pi)^{3(n-1)} N_I!} \prod_{l=1}^n (2s_l+1),
\end{equation}
where $x$ is the invariant mass of the fireball, $P_n^d$ is the probability that one get $n=2$ or $n=3$ hadrons from one fireball, $\Phi_n(E,m_1,..,m_n)$ is the n-body phase space integral, consisting the Breit-Wigner factors for resonant particles, $\rho(E)$ is the density of states obtained from the statistical Bootstrap, $N_I$ is a symmetry factor, and $s_l$ is the total spin of the $l'th$ particle. All of these factors are described in details in \cite{Main1,Main2,Main3}.

The final ingredient to the hadronization probability is the quark combinatorial probability factor $C_{Q}(x)$, which has to be modified to include tetraquarks into the model, therefore it will be given a more detailed discussion here. 

This factor basically counts all the possible combinations of how the quarks and antiquarks could build up specific hadrons, taking into consideration a color factor, and a statistical factor for the same type of quarks, then normalized by the possible final states for the $n=2$ and $n=3$ body cases. The total number of quarks and antiquarks are given by simple phase space considerations \cite{CHIPS}. Assuming massless quarks the n-body phase space can be expressed as:

\begin{eqnarray}
\label{eq:extra1}
\nonumber
&{\displaystyle \Phi_n^{\text{quarks}}(x) = \int \prod_{i=1}^{n} \frac{d^3p_i}{2E_i(2\pi)^3}(2\pi)^4\delta^4(P_{\mu}-\sum_{i=1}^{n}p_{\mu,i}) = } \\
& {\displaystyle \frac{1}{2(4\pi)^{2n-3}}\frac{x^{2n-4}}{\Gamma(n)\Gamma(n-1)} },
\end{eqnarray}
where $\Gamma(n)$ is the gamma function, $x$ is the energy, $p_i$ represents three-momenta, while $p_{\mu,i}$-s are four-momenta. Using an $e^{-x/T_0}$ statistical weight factor with a distribution proportional to $\Phi_n(x)$ the mean value of $\langle x^2 \rangle$ at temperature $T_0$ can be expressed as:
\begin{equation}
\label{eq:extra2}
\langle x^2 \rangle = \frac{ \int dx x^2 \Phi_n^{\text{quarks}}(x) e^{-x/T_0}}{\int dx \Phi_n^{\text{quarks}}(x)e^{-x/T_0} }
\end{equation}
The expression for $\langle x^2 \rangle$ will depend on the number of quarks $N$ as:
\begin{equation}
\label{eq:extra3}
\langle x^2 \rangle = 4 N(N-1)T_0^2
\end{equation}
Solving Eq.\ref{eq:extra3} for $N$, the energy dependence of the number of constituents can be given as:
\begin{equation}
N(x) = \frac{1+\sqrt{1+x^2/T_0^2}}{2}
\label{eq:30}
\end{equation}
where $x$ is the energy, and $T_0$ is a free parameter set to $T_0=160$ MeV. At this point there is no distinction between quark flavours as the calculation is made to massless quarks, which is a good approximation for the up and down quarks, however for the strange, and especially the heavier bottom, charm and top quarks it is not straightforward, how to include them into the picture. In our model, we define a - quark creational probability - for each quark, which gives the probability that a specific type of quark is created. This is a free parameter of the model, however reasonable restrictions can be made e.g. $P_u=P_d$, which was enough for calculating low energy exclusive cross section ratios, containing only up and down quarks. To calculate more interesting processes e.g. processes including heavy quarks the fit for $P_s,P_c,P_b$ were also necessary. \footnote{We do not consider top quarks as the energy range, we intend to use the model is small compared to their masses.}

Knowing the quark creational probabilities the probability of a specific quark (antiquark) number configuration can be given by a multinomial distribution as:
\begin{equation}
\label{eq:31}
F(x,n_i) = \frac{N(x)!}{\prod_{i}n_i!} \prod_{i} P_i^{n_i},
\end{equation}
where $n_i$ is the number of quarks (antiquarks) of type $i=u,d,s,c,b$ satisfying the constraint $N=\sum_i n_i$, and $P_i^{n_i}$ is the quark creational probability, which gives the probability that a quark is of the type $i$. The expected number of quarks of type $i$ corresponds to the maximum of the distribution function if $n_i \ge 1$ and is given by $\langle n_i(x) \rangle = P_i N(x)$. For small quark creational probabilities (charm and bottom quarks at low energies) the distribution is constrained by assuming that $n_i=1$ with taking the maximum of the constrained probability. This will give the same suppression for these quarks as what we would get from simply taking the expectation values, so the following two methods are equivalent: (1) taking the expected number of quarks (2) taking the quark numbers corresponding to the maximum or the constrained maximum of the distribution.

The model is used at relatively low energies (few GeV's) up to a few tens of GeV's where the quark creational probabilities are fitted through measured exclusive and inclusive cross sections \cite{Main3}. At these energies a constant value for the up, down and strange quarks were satisfactory, however for the heavier quarks (charm,bottom) a simple linearly rising probability is assumed as $P_{c,b}=a_{c,b}x$. At higher energies a functional form with a saturation is a better option, however this simple linear approximation is good at the energies, we intend to use the model. The fitted parameters are $P_u=P_d=0.425$, $P_s=0.15$, $a_c = 8.5  \cdot 10^{-4}$ GeV$^{-1}$, and $a_b = 1.05 \cdot 10^{-5}$ GeV$^{-1}$. The smaller values for the strange, charm and bottom quarks are connected to their heavy masses and suppression factors, which are usually used in statistical models \cite{Strange}. Also it is worth to note, that due to the running nature of the charm and bottom quark creational probbilities the other parameters have to run as well, however because the charm and bottom probabilities are so small, the values shown here can be used here. The detailed explanation for these parameters and their validity is given in \cite{Main3}. 
To validate the model for tetraquark production it will be necessary to know the quark creational probabilities at $\sqrt{s}=7$ TeV, which is clearly out of the range of our previous fits. It is also really hard to find suitable processes, which are well measured at this energy and are comparable to make an estimation to the quark creational probabilities, however a sensible estimation can be made using the previous reasoning for the saturating probabilities. This will be described in details in Sec.\ref{sec:3}.
Moving forward, from the known number of quarks the combinatorial factor is given by counting all the possible combinations of quarks, giving the desired final state. This factor is then multiplied by a color factor, which counts the possible number of colorless combinations for the mesons and baryons.
After normalization the quark combinatorial probability $C_{Q}(x)$ is obtained, which is of course energy dependent as the number of quarks also depend on the energy.
In this model the quark content information is hidden in the quark combinatorial factors, therefore to include diquarks it has to be modified. We use the simplest approximation possible and assume that the probability of forming a diquark is simply given as $P_{ij} = P_{i}P_{j}$. Including the diquarks into the distribution in Eq.\ref{eq:31} we get:
\begin{equation}
\label{eq:32}
F(x,n_i,n_{ij}) = \frac{N(x)!}{\prod_{i}n_i! \prod_{ij}n_{ij}!} \prod_{i} P_i^{n_i}\prod_{ij} P_{ij}^{n_{ij}},
\end{equation}
with the same constraint that the sum of the number of specific type of quarks has to give back the total number of quarks. For the diquarks $n_{ij}$ means that one take $n_{ij}$ number of $i$ quarks and $j$ quarks seperately, so the maximum number of diquarks is of course $N/2$ and the expected number is $\langle n_{ij} \rangle = P_i P_j N/2$. 

The next difference between diquark picture and the original quark picture is the number of color configurations, where in the original case this factor simply counts the number of colorless configurations for the mesons and baryons built up from color triplet quarks, however the diquarks could be in a color triplet or sextet state as well. 

The possible states built from the quarks are the color antitriplet $\frac{1}{\sqrt{2}}(rg-gr)$, $\frac{1}{\sqrt{2}}(rb-br)$, $\frac{1}{\sqrt{2}}(gb-bg)$, and the color sextet $rr$, $gg$, $bb$, $\frac{1}{\sqrt{2}}(rg+gr)$, $\frac{1}{\sqrt{2}}(rb+br)$, $\frac{1}{\sqrt{2}}(gb+bg)$. Similiarly for the antiquarks the possible states are the triplet $\frac{1}{\sqrt{2}}(\overline{r}\overline{g}-\overline{g}\overline{r})$, $\frac{1}{\sqrt{2}}(\overline{r}\overline{b}-\overline{b}\overline{r})$, $\frac{1}{\sqrt{2}}(\overline{g}\overline{b}-\overline{b}\overline{g})$, and antisextet  $\overline{r}\overline{r}$, $\overline{g}\overline{g}$, $\overline{b}\overline{b}$, $-\frac{1}{\sqrt{2}}(\overline{r}\overline{g}+\overline{g}\overline{r})$, $\frac{1}{\sqrt{2}}(\overline{r}\overline{b}+\overline{b}\overline{r})$, $-\frac{1}{\sqrt{2}}(\overline{g}\overline{b}+\overline{b}\overline{g})$. The number of states therefore which could build up a singlet state from a diquark and an antidiquark in the triplet-antitriplet and sextet-antisextet configuration are 3 and 6 respectively. If the diquark consist quarks of the same kind e.g. (uu), then an extra $1/2$ factor appears because of indistinguishability. Lastly, the diquark could be in a specific spin configuration and therefore an extra total spin factor arises for the bound diquarks as well. In Sec.\ref{sec:4} an example is given for the $X(3872)$ particle on how to calculate its cross section by the statistical model described here.

\section{Estimating the quark creational probabilities at high energies}
\label{sec:3}
As it was mentioned before the quark creational probabilities are free parameters of the model and they have to be fitted from experiments or estimate them from some theoretical model. For the light up and down quarks the simple constraint $P_u=P_d$ seemed reliable even at low energies, however for the heavier strange quark a smaller $P_s$ value had to be applied, which was fitted through Kaon production cross sections. For the charm and bottom quarks $P_c$ and $P_b$ were fitted through inclusive chamonium and bottomonium cross sections up to a few tens of GeV's. For higher energies $\sqrt{s} \ge 100$ GeV a fit were more problematic due to numerical complexities and sparse measured data, which could be used to fit $P_c$ and $P_b$. From the fits it was clear that the suppression mainly depends on the mass of the quarks (the higher the mass, the larger the suppression), and the suppression becomes smaller as the energy gets higher. For the strange quarks the creational probability could be described by a constant factor in the energy range from a few GeV up to $15$ GeV, however it is worth mentioning that a slowly linearly rising $P_s$ could also be fitted instead of a constant value, giving slightly better results for strangeness production, however for simplicity, we still used a constant value in this energy range, because the model errors were larger than the difference caused by this change. At higher energies the linearly rising $P_s$ should be used instead. 

As the only measured $X(3872)$ cross section data is at $\sqrt{s}=7$ TeV, we should give an estimation to the quark creational probabilities at this energy, to be able to validate the model. For the light quarks (u,d,s) it would be safe to assume that at $7$ TeV $P_u=P_d=P_s$. For the heavier charm, bottom and top quarks this is not that evident. For the top quark due to its very high mass, we could assume that $P_t \ll P_{u,d,s}$ still can be applied at this energy, so we are left only with the charm and bottom quarks. We will approximate their relative probabilities with using two models. 

For the first estimation, we will use the fitted functional form of the charm and bottom creational probabilities and calculate the energies, when $P_c$ and $P_b$ will be approximately equal to $P_u$,$P_d$, and $P_s$, while of course satisfiyng the $P_u + P_d + P_s + P_c + P_b +P_t= 1$ constraint, where $P_t \ll 1$ is assumed to be negligible. To do this, we will extrapolate the $P_{c,b}=a_{c,b}E$ linear approximation to energies in the TeV range. As the charm quark creational probability will reach the desired value faster, the first equation to be solved is:
\begin{equation}
3P_{uds} = 1 - P_c -P_b
\end{equation}
where $P_{uds}\equiv P_u=P_d=P_s$, while $P_c=a_c E$, and $P_b=a_b E$, with the fitted values $a_c=8.5 \cdot 10^{-4}$ GeV$^{-1}$,  $a_b=1.05 \cdot 10^{-5}$ GeV$^{-1}$. The energy when the charm quark creational probability will reach $P_{uds}$ can be calculated if we substitute $P_{uds}$, with $P_c$, which gives the energy:
\begin{equation}
E_c = \frac{1}{4a_c+a_b} \approx 300 GeV
\end{equation}
This means that according to the extrapolated fit, the charm quark crational probability will be close to $P_u$, $P_d$, and $P_s$ at $\sqrt{s} \approx 300$ GeV. Following the same procedure for the bottom quarks the energy, where it will be close to $P_{u}=P_d=P_s=P_c$ can be expressed as:
\begin{equation}
E_b=\frac{1}{5a_b}\approx 20 TeV
\end{equation}
These results suggest that at $\sqrt{s}=7$ TeV, the bottom quark creational probability will still be suppressed, however the charm quark creational probability will be close to the up, down, and strange quark creational probabilities. The exact values for the suppression are not really relevant, as we used an extrapolation of the low energy fit at much higher energies and it is possible that the simple linear fit is not valid at close to equal probabilities. This is also suggested from the strange quark creational probabilities at lower energies, where the rising of $P_s$ was almost constant near $P_u$, and $P_d$, which could mean that the rising of the probabilities will be slowing down as the energy goes higher. However, at this point, we only want to estimate the energies, where the probabilities will be close together and we do not care about their actual functional form.

\begin{figure}
\resizebox{0.5\textwidth}{!}{%
  \includegraphics{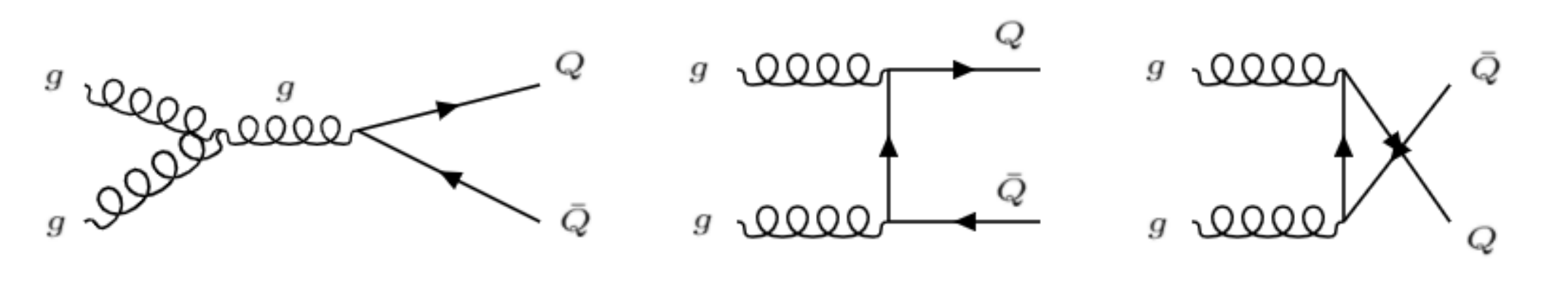}
}
\caption{Feynman diagrams for heavy quark pair production from gluon fields.}
\label{fig:Feynman}       
\end{figure}

In the next model an estimation for the possible suppression is given. In this model, we assume that the creation of heavy quark pairs from gluon fields can be described perturbatively \cite{extra1}, and calculate the cross section ratios of the charm-anticharm, and bottom-antibottom pair productions. In this simplified model, we assume that the quark-antiquark pairs are generated from the gluon fields, through the first order perturbative processes shown in Fig.\ref{fig:Feynman}. From this model, we excluded some other interesting processes like $q \bar{q} \rightarrow Q \bar{Q}$, gluon splitting, and other higher order processes as well for simplicity. 

The total cross section for the process $p p \rightarrow Q \bar{Q}$ can be calculated using the factorization theorem \cite{extra2}:
\begin{eqnarray}
\label{eq:extra4}
\nonumber
&{\displaystyle \sigma(\sqrt{s})^{pp \rightarrow Q \bar{Q}} = \int dx_1 dx_2 f_{1,p}(x_1,\mu_F)f_{2,p}(x_2,\mu_F)  \times } \\
&{\displaystyle  \times \hat{\sigma}^{gg\rightarrow Q \bar{Q}}(\hat{s},\mu_R)}
\end{eqnarray}
where  $\mu_F$ is the factorization scale, $\mu_R$ is the renormalization scale, $x_1$, and $x_2$ are the momentum fractions, $f_{i,p}$ is the proton structure function, and $\hat{s}=x_1x_2s$. It is common to set the factorization and the renormalization scale to the mass of the heavy quark, so $\mu_F=\mu_R=m_Q$. Furthermore, $\hat{\sigma}^{gg\rightarrow Q\bar{Q}}$ is the elementary gluon-gluon cross section described by Fig.\ref{fig:Feynman}, which has the form:
\begin{eqnarray}
\label{eq:extra5}
\nonumber
&{\displaystyle\hat{\sigma}^{gg\rightarrow Q \bar{Q}} = \frac{\alpha_s^2}{m_Q^2}\frac{\pi \beta \rho}{24 N (N^2-1)} [
3 \eta(\beta)(\rho^2+
 } \\
\nonumber
&{\displaystyle  +2(N^2-1)(\rho+1))  + 2(N^2-3)(1+\rho) + } \\
&{\displaystyle  + \rho(6\rho-N^2)]}
\end{eqnarray}
where
\begin{eqnarray}
\label{eq:extra6}
\nonumber
&{\displaystyle  \eta(\beta)=\frac{1}{\beta}\log \Big( \frac{1+\beta}{1-\beta}\Big)} \\
&{\displaystyle \beta = \sqrt{1-\rho}} \\
\nonumber
&{\displaystyle \rho = \frac{4m_Q^2}{\hat{s}}}
\end{eqnarray}
where $m_Q$ is the mass of the charm, or bottom quarks, $\hat{s}=x_1x_2s$ is the partonic center of mass energy, while $\alpha_s$ is the strong coupling constant, defined by:
\begin{equation}
\alpha_s(Q^2) = \frac{12 \pi}{(33-2n_f)\ln \Big( \frac{Q^2}{\Lambda_{QCD}^2}\Big)}
\end{equation}
where $n_f$ is the number of quark flavours, $Q$ is the scale of the process, set by the minimum virtuality exchanged in the t-channel diagram in Fig.\ref{fig:Feynman} to $Q = m_Q$, while $\Lambda_{QCD}$ is set to $0.2$ GeV.

\begin{figure}
\resizebox{0.5\textwidth}{!}{%
  \includegraphics{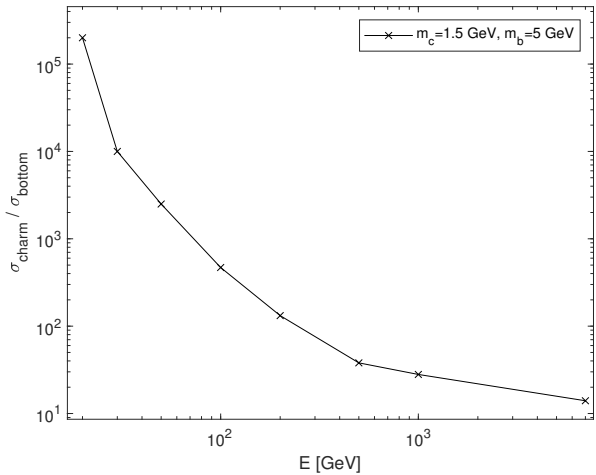}
}
\caption{Ratio of the charm-anticharm and bottom-antibottom pair production cross sections in proton-proton collisions.}
\label{fig:charm_bottom}       
\end{figure}

To calculate the hadronic cross sections the parton distribution functions (pdf's) are also needed \cite{CTEQ1,CTEQ2}. In Fig.\ref{fig:charm_bottom} the ratio of the charm and bottom production cross sections can be seen up to $\sqrt{s}=7$ TeV, using the values $m_c=1.5$ GeV, and $m_b=5$ GeV. It can be seen that the bottom quark production rate is slowly getting closer to the charm production rate, however at 7 TeV it is still an order of magnitude smaller.

Our main conclusion in this section is that at $\sqrt{s}=7$ TeV the charm quark creational probability will be close to the up,down, and strange quark creational probabilities, however the bottom quark creational probability will be approximately an order of magnitude smaller than the previous ones. Therefore in the followings, we set these values to:
\begin{equation}
P_u^{7TeV}=P_d^{7TeV}=P_s^{7TeV}=P_c^{7TeV} \approx 0.2439
\end{equation}
\begin{equation}
P_b^{7TeV} \approx 0.0244
\end{equation}

\section{Results}
\label{sec:4}
In this section the statistical model described in Sec.\ref{sec:2} is used to estimate the measured high energy cross section of the $X(3872)$ particle at $7$ TeV and to make predictions to its low-energy cross sections in proton-proton, proton-antiproton, and pion-proton collisions. 

The inclusive $X(3872)$ production cross section in proton-proton collisions at $\sqrt{s} = 7$ TeV is measured by the CMS collaboration using the decays $X(3872) \rightarrow J/\Psi \pi^+\pi^-$ and $\Psi(2S)\rightarrow J/\Psi \pi^+\pi^-$ with the subsequent dilepton decay of the charmonium $J/\Psi \rightarrow \mu^+ \mu^-$ , where the ratio of the two channels in the kinematical region of $p_T \in [10,50]$ GeV, and $|y|<1.2$ are measured to be \cite{X_MEAS1}:
\begin{equation}
\label{eq:33}
\frac{\sigma_{X(3872)} \cdot \text{Br}(J/\Psi \pi^+ \pi^-)}{\sigma_{\Psi(2S)}\cdot \text{ Br}(J/\Psi \pi^+ \pi^-)} = 0.0656 \pm 0.0094
\end{equation}
where  the error contains the systematical and statistical errors as well. The branching factor for the charmonium is well measured $\text{ Br}(\Psi (2S) \rightarrow J/\Psi \pi^+ \pi^-) = 0.34$, however for the $X(3872)$ a lower and upper bound is given in \cite{X_MEAS2} $ \text{Br}(X(3872) \rightarrow J/\Psi \pi^+ \pi^-) = [0.042,0.093]$. Using the branching fractions, while neglecting the measurement error the ratio interval of the cross sections is given as:
\begin{equation}
\label{eq:34}
\frac{\sigma_{\Psi (2S)}}{\sigma_{X(3872)}} \approx [1.88,4.16]
\end{equation}
We use this interval to validate our assumptions for the tetraquark model, where we also calculated the inclusive cross section ratio at $\sqrt{s}= 7$ TeV. This energy is clearly out of the energy range, we previously used our model, and the values for the previously fitted quark creational probabilities has to be readjusted, especially for the heavier quarks, where it can be assumed that at these high energies the functional form of the charm and bottom quarks should be close to a plateau. In Sec.\ref{sec:3}, we made some estimation for the quark creational probabilities at $\sqrt{s}=7$ TeV, which estimation will be used here. By taking the ratios of the corresponding cross sections many of the common factors are simply drop out and we will have a simple closed form for the ratio.

\begin{figure}
\resizebox{0.5\textwidth}{!}{%
  \includegraphics{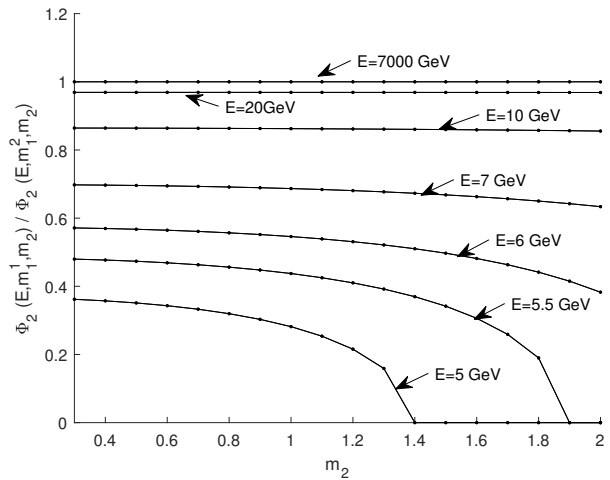}
}
\caption{Ratio of the two-body phase spaces with different masses at different energies. At very high energies at the TeV scale the phase spaces are almost equal.}
\label{fig:ratio}       
\end{figure}

We assume that the tetraquark is a $J^{PC}=1^{++}$ state and has the quark content of $(u\overline{u}c\overline{c}$). If the diquarks are in the s-state, then a possible spin configuration, which can give the desired parity, spin, charge conjugation and satisfy the Pauli principle is when one diquark has spin 1 and the other has spin 0 \cite{Tetraquark_SPIN,TetraquarkDiquarkSpin}. It is straightforward to check that the only factors, which are not dropped out of the ratio are the quark combinatorial factors and the phase space integrals. Due to the almost similiar masses of the $\Psi(2S)$ and $X(3872)$ states, we assume that the phase spaces do not make too much difference at $7$ TeV, therefore:
\begin{equation}
\Phi_2(x,m_{X(3872)},m_i)\approx\Phi_2(x,m_{\Psi(2S)},m_i)
\label{eq:35}
\end{equation}
\begin{equation}
\Phi_3(x,m_{X(3872)},m_i,m_j)\approx\Phi_3(x,m_{\Psi(2S)},m_i,m_j)
\label{eq:36}
\end{equation}
at $x=7$ TeV. To show that this assumption is valid, we calculated the two-body phase space integrals $\Phi_2(E,m_1^i,m_2)$ for two different cases with different fixed masses $m_1^{i=1}=3.6$ GeV, and $m_1^{i=2}=1$ GeV, and varied the energies and the mass of the second particle between, $E \in [5,7000]$ GeV, and $m_2 \in [0.3,2]$ GeV. In this way the masses of the first particles $m_1^i$ corresponds to the masses of the particles of interest, while $m_2$ serves as a dummy-variable to check the difference that the mass of the second particle could make in the ratio $\Phi_2(E,m_1^{i=1,}m_2)/\Phi_2(E,m_1^{i=2},m_2)$ at a specific energy. The results can be seen in Fig.\ref{fig:ratio}, where it can be deduced that while at low energies there could be a significant difference in the phase spaces, especially when the mass of the second particle is higher, however at very high energies the differences in the phase spaces are gone, and we can safely use the mentioned assumption. The arguments made for the two-body phase space are valid even for the three- and more-body phase spaces as well.

\begin{figure}
\resizebox{0.5\textwidth}{!}{%
  \includegraphics{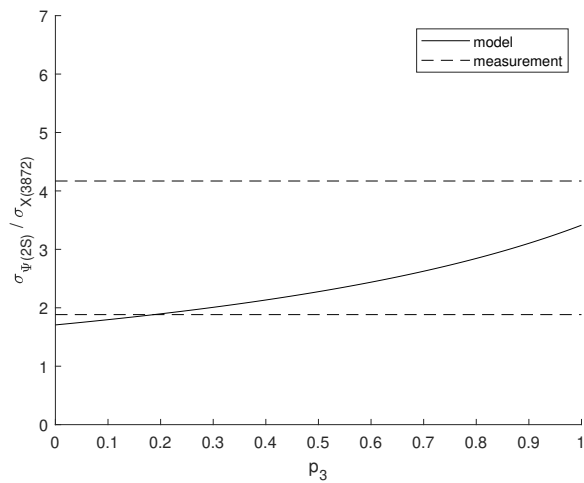}
}
\caption{Ratio of the $\Psi(2S)$ and $X(3872)$ production cross sections in proton-proton collisions at $\sqrt{s}=7$ TeV. The dashed lines correspond to measurements taking into consideration the upper and lower limit of the $X(3872) \rightarrow J/\Psi \pi^+ \pi^-$ decay width, while the full black line is the calculated ratio from the statistical model for different $p_3$ triplet-antitriplet probabilities. }
\label{fig:6}       
\end{figure}

Using this assumption the quark combinatorial factors can be taken out from the one fireball inclusive sums and the ratio can be expressed by using only the quark combinatorial factors as:
\begin{eqnarray}
\label{eq:37}
&{\displaystyle r^{\Psi(2S)}_{ X} \approx \frac{ ( \langle n_c\rangle^0 - \sum_{i=u,d,s,b} \langle n_{ic}\rangle  - 2\langle n_{cc}\rangle )^2}{3 p_3 \langle n_{uc}\rangle^2 + 6(1-p_3) \langle n_{uc}\rangle^2 }  } = \\ \nonumber
& {\displaystyle = \frac{\Big(P_c - \sum_i \frac{P_i P_c}{2} - P_c^2\Big)^2}{P_u^2 P_c^2 \Big( \frac{3}{4}p_3 + \frac{6}{4}(1-p_3) \Big)}}
\end{eqnarray}
where $\langle n_i \rangle^0 = P_i N$, and  $\langle n_{ij} \rangle = P_i P_j N/2$ are the expected number of quarks/diquarks of the type $i=u,d,s,c,b$, while the $p_3$ factor is the probability of the triplet combination. The common total spin, phase space, etc... factors are dropped out and for the charmonium state the $  \langle n_c\rangle^0 - \sum_{i=u,d,s,b} \langle n_{ic}\rangle  - 2\langle n_{cc}\rangle $ factor is the expected number of the charm quarks after the charm quarks contained by the possible diquarks are subtracted from the total number of charm quarks. As at this point it is not know if the tetraquark is in the triplet-antitriplet or the sextet-antisextet diquark-antidiquark state, we checked the results for the full interval $p_3 \in [0,1]$ which can be seen in Fig.\ref{fig:6}.
The results we obtained are in the desired range for almost every value of $p_3$, which suggest that its actual value is not that important, at least in our model with the current measurement uncertainties considered. In the followings, we will assume that $p_3=1$, thus the tetraquark is in the triplet-antitriplet state.

\begin{figure}
\resizebox{0.5\textwidth}{!}{%
  \includegraphics{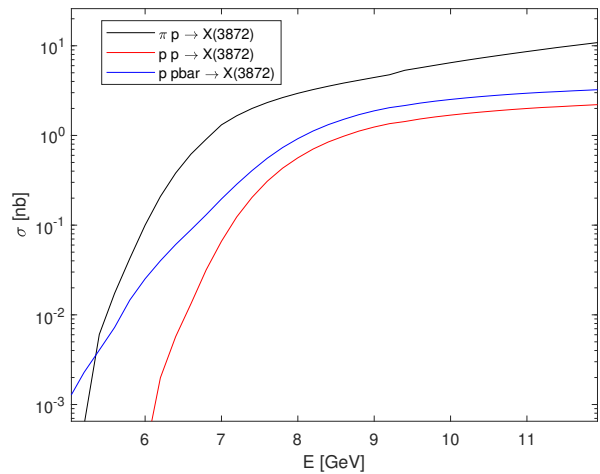}
}
\caption{Inclusive production cross sections of the $X(3872)$ state assuming it is a diquark-antidiquark bound state. Only the direct production (without the $B \rightarrow X(3872)K$ decays) is considered.}
\label{fig:7}       
\end{figure}

To estimate the energy dependent inclusive cross sections at lower energies, the inclusive sums are calculated in the two fireball scheme, as it has been done for e.g. charmonium states in \cite{Main3}. Here, the previously fitted low energy quark creational probabilities, and not the ones estimated at $7$ TeV are used. Also the phase space is calculated correctly for the tetraquark as well, without using the assumptions in Eq.\ref{eq:35} and Eq.\ref{eq:36}. The results without the model errors for direct $X(3872)$ tetraquark production in proton-proton, pion-proton, and proton-antiproton collisions can be seen in Fig.\ref{fig:7}.

\section{Conclusions}
\label{sec:5}
In this paper, we used a statistical based model to estimate tetraquark production cross sections, namely the $X(3872)$ possible four-quark state in proton-proton, pion-proton and proton-antiproton collisions from a few GeV center of mass energies up to $12$ GeV. To validate the statistical model the inclusive cross section ratio for the $\Psi(2S)$ and $X(3872)$ particles are calculated and compared to the available measured value at $\sqrt{s}=7$ TeV in proton-proton collisions. In the extension of the statistical model the only new free parameter is the diquark-antidiquark color configuration probability $p_3$, however the results are satisfactory for almost all of its values, and it could be assumed that the triplet-antitriplet configuration is more likely than the sextet-antisextet one. Due to the large uncertainty in the measurements the $p_3$ probaiblity cannot be fitted, therefore, we assume that the tetraquark is in the triplet-antitriplet state. Using the extended model, estimations were made to the inclusive cross sections of the $X(3872)$ states in proton-proton, pion-proton and proton-antiproton collisions giving a smaller production cross section as one will get for a simple charmonium states, which behaviour is expected. These cross sections will be useful in some transport calculations \cite{Transport1,Transport2,Transport3}, we intend to do in the future, where a distinction could be made in the structure of the $X(3872)$ if it is a compact tetraquark-, or a loosely bound molecule state, because they will have different production and dissociation cross sections. Calculations similiar to these have already been done \cite{X_TRANSPORT1,X_TRANSPORT2,X_TRANSPORT3,X_TRANSPORT4,X_TRANSPORT5}, and it would be interesting to compare the results from different transport codes, and from different approaches. The necessary elementary cross sections for the tetraquark state in the diquark picture could be given using the model described here.

%

%

\bibliographystyle{unsrt}
 \bibliography{Main_Document}

%
%
%


\end{document}